\documentclass[conf]{new-aiaa}
\usepackage[utf8]{inputenc}

\usepackage{graphicx}
\usepackage{amsmath}
\usepackage[version=4]{mhchem}
\usepackage{siunitx}
\usepackage{subcaption}
\usepackage{adjustbox}
\usepackage{longtable,tabularx}
\usepackage{booktabs}
\usepackage{blkarray}

\title{Gradient‑Informed Monte Carlo Fine‑Tuning of Diffusion Models for Low‑Thrust Trajectory Design}

\author{Jannik Graebner\footnote{Ph.D.\ Student, Department of Mechanical and Aerospace Engineering,
Princeton University} and Ryne Beeson \footnote{Assistant Professor, Department of Mechanical and Aerospace Engineering,
Princeton University.}}
\affil{Princeton University, Princeton, NJ, 08544}

\newcommand{\pd}[2]{\frac{\partial #1}{\partial #2}}
\DeclareMathOperator*{\argmin}{arg\,min}

\begin{document}

\maketitle

\begin{abstract}
Preliminary mission design of low‑thrust spacecraft trajectories in the Circular Restricted Three‑Body Problem is a global search characterized by a complex objective landscape and numerous local minima.
Formulating the problem as sampling from an unnormalized distribution supported on neighborhoods of locally optimal solutions, provides the opportunity to deploy Markov chain Monte Carlo methods and generative machine learning.
In this work, we extend our previous self-supervised diffusion model fine-tuning framework to employ gradient‑informed Markov chain Monte Carlo.
We compare two algorithms - the Metropolis‑Adjusted Langevin Algorithm and Hamiltonian Monte Carlo - both initialized from a distribution learned by a diffusion model.
Derivatives of an objective function that balances fuel consumption, time of flight and  constraint violations are computed analytically using state transition matrices.
We show that incorporating the gradient drift term accelerates mixing and improves convergence of the Markov chain for a multi-revolution transfer in the Saturn-Titan system.
Among the evaluated methods, MALA provides the best trade-off between performance and computational cost. Starting from samples generated by a baseline diffusion model trained on a related transfer, MALA explicitly targets Pareto-optimal solutions. Compared to a random walk Metropolis algorithm, it increases the feasibility rate from $17.34\%$ to $63.01\%$ and produces a denser, more diverse coverage of the Pareto front.
By fine-tuning a diffusion model on the generated samples and associated reward values with reward-weighted likelihood maximization, we learn the global solution structure of the problem and eliminate the need for a tedious separate data generation phase.
\end{abstract}

\section{Introduction}
Space missions with electric propulsion demand trajectories that simultaneously minimize fuel consumption and time of flight.
Identifying Pareto-optimal solutions with respect to these objectives for long-duration, low-thrust spacecraft trajectories is challenging and computationally expensive. 
A high-dimensional solution space, coupled with the nonlinear and non-convex dynamics of the Circular Restricted Three-Body Problem (CR3BP) leads to an optimal control problem with a complex objective landscape and numerous local minima. 

Applying an indirect optimal control approach reduces the dimension of the solution space through the introduction of costate variables, which become the decision variables of the problem. 
However, the lack of physical intuition for the costates and the high sensitivity of the solution to their values make the practical use of this method challenging. 
Efficient performance therefore depends on generating good initial costate guesses.
Previous work has shown that locally optimal solutions for trajectory optimization problems tend to form clusters \cite{Yam.2011, Li.872023, Graebner.1032024}.
These structures can be leveraged when constructing high-quality initial costate guesses.
By defining a probability density function supported on solution clusters, the global search problem is recast as sampling from a distribution with unnormalized density. 
Using Monte Carlo methods and generative machine learning, we combine two popular families of sampling methods, leveraging the complementary strengths of both approaches.

In this work, we extend our previous framework, which fine-tunes diffusion models using samples from a Markov chain Monte Carlo (MCMC) algorithm to learn a global solution distribution for an indirect spacecraft trajectory optimization problem \cite{graebner2025aas}. 
The novel contribution is to incorporate gradient information into the MCMC algorithm, which leads to faster convergence of the Markov chain. 
We employ two gradient-based MCMC algorithms: the Metropolis-Adjusted Langevin Algorithm (MALA) \cite{Roberts.1998} and Hamiltonian Monte Carlo (HMC) \cite{Neal.2012}. 
The performance of both algorithms is benchmarked against the random-walk Metropolis algorithm (RWM) used in the previous framework. 
In both cases, we introduce a new formulation of the objective function, which is efficiently evaluated through a preliminary screening algorithm. 
We show that the gradient of this reformulated objective function can be reasonably approximated through a fixed-time objective.
This approximated gradient is then computed analytically through the propagation of state transition matrices (STMs). 
While the gradient computations make each iteration of the MCMC algorithm more expensive, the gradient-based methods require fewer iterations to generate high-quality samples. 
We test the framework on a planar, multi-revolution transfer in the Saturn–Titan system and show that MALA achieves the highest sample quality for comparable runtimes across all three algorithms. 
We then demonstrate how the MALA algorithm is incorporated into the self-supervised fine-tuning framework to learn a global solution distribution without a separate, tedious data generation phase. 
The improved efficiency of the framework highlights its potential for application as part of preliminary mission design.

\section{Related Work}
The idea of exploiting the clustering structure of locally optimal solutions has long been employed in global search algorithms such as Monotonic Basin Hopping (MBH) \cite{Wales.1997}. 
However, MBH relies on simple sample distributions and requires extensive manual parameter tuning. 
Li et al. \cite{Li.872023} were the first to leverage generative models, specifically conditional variational autoencoders and diffusion models, to create informed initial guesses for trajectory optimization. 
Trained on databases of pre-computed solutions across multiple thrust levels, their models warm-started a direct optimization solver for unseen thrust values, substantially improving convergence rates \cite{Li.872023,Li.2222024,Graebner.1032024}.

Extending these ideas, we recently proposed a global-search framework that pairs diffusion models with the indirect optimal-control method for low-thrust trajectory design \cite{Graebner.1132025}. 
Diffusion models are state-of-the-art generative machine learning models that first gained popularity in the field of image generation \cite{SohlDickstein.3122015,Ho.6192020}. 
In the context of indirect spacecraft trajectory optimization, they are employed to generate high-quality initial costates. 
Given a training dataset consisting of costate–control solutions for a specific transfer with varying values of a mission parameter, the model is trained to learn the underlying data distribution conditioned on that parameter. 
In the sampling stage, the model can then generate new solutions from the learned distribution for given mission parameters.
\begin{figure}[b!]
\centering
\includegraphics[width=.9\textwidth]{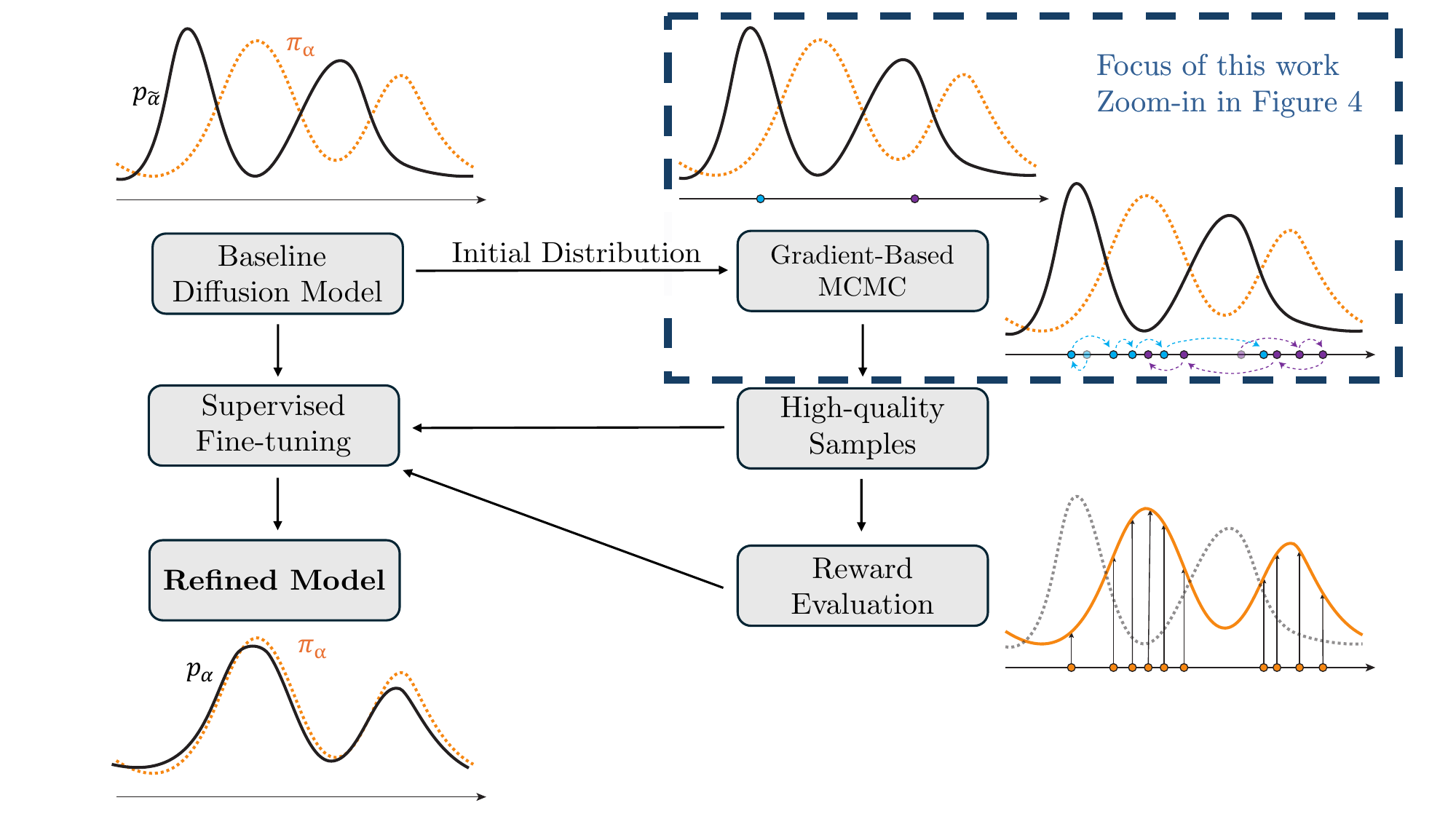}
\caption{Simplified illustration of the sampling framework: starting from a baseline diffusion model with distribution $p_{\tilde{\alpha}}$ that is not aligned with the target distribution $\pi_\alpha$, gradient-based MCMC and supervised fine-tuning are used to train a refined model that closely matches the target distribution.}
\label{fig: MCMC_framework}
\end{figure}

Although this framework produces high-quality, diverse samples that closely mirror the true solution distribution, it still incurs two key drawbacks: (i) the high up-front cost of generating labeled training data and (ii) limited generalization when the transfer scenario changes. 
We addressed both of these issues by embedding the diffusion model inside a self-supervised MCMC loop \cite{graebner2025aas}. 
An overview of the updated version of this framework with gradient-based MCMC extensions is shown in Figure~\ref{fig: MCMC_framework}.
Starting with samples from a baseline diffusion model trained on a related transfer, a gradient-based MCMC algorithm is employed to generate high-quality samples from a target distribution.
These samples, together with their respective reward values, are then used to fine-tune the baseline diffusion model, leading to an improved model that achieves a closer fit to the target distribution.

\section{Problem Formulation}
The following setup is similar to our previous work  \cite{Graebner.1132025} and we restate the problem setup for the convenience of the reader.

\subsection{Optimal Control Problem}
\label{sec: optimal control problem}
Optimal control aims to determine a feasible control history $\boldsymbol{u}(t)$ for a dynamical system that minimizes a specified performance measure while satisfying state constraints \cite{Kirk.2004}.
A general version of the problem is given by
\begin{align}
    \label{equation: cost functional}
    \min J(\boldsymbol{u}) \equiv \phi(\boldsymbol{x}(t_f), t_f) + \int_{t_0}^{t_f} \mathcal{L}(\boldsymbol{x}(t), \boldsymbol{u}(t), t) dt \quad 
    \textrm{subj. to Eqs. } 
    \eqref{equation: evolution differential equation}, \eqref{equation: initial and final boundary conditions}, \eqref{equation: path constraints} \ , 
\end{align}
where $J$ is the objective function which includes a running cost $\mathcal{L}$ and a terminal cost $\phi$.
We consider a free final time problem over the interval $t\in[t_0,t_f]$.
The state $\boldsymbol{x}(t)$ evolves according to the dynamics of the system
\begin{align}
    \label{equation: evolution differential equation}
    \dot{\boldsymbol{x}}(t) = \boldsymbol{f}(\boldsymbol{x}(t), \boldsymbol{u}(t), t), \quad \forall t \in [t_0, t_f],
\end{align}
and satisfies the initial and terminal boundary conditions
\begin{align}
    \label{equation: initial and final boundary conditions}
    \boldsymbol{x}(t_0) = \boldsymbol{x}_0, \quad  \boldsymbol{\psi}\left[ \boldsymbol{x}(t_f), t_f \right] = \boldsymbol{0}.
\end{align}
Both the natural dynamics and control-induced accelerations are included in the vector field $\boldsymbol{f}$.
Additional constraints on the state and control are prescribed as equality path constraints
\begin{align}
    \label{equation: path constraints}
    \boldsymbol{\xi}(\boldsymbol{x}(t), \boldsymbol{u}(t), t) = \boldsymbol{0}, \quad & \forall t \in [t_0, t_f],
\end{align}
with no inequality path constraints considered.

\subsection{Low Thrust Spacecraft Trajectory Optimization}
Our goal is to find Pareto-optimal trajectories for a low-thrust transfer, minimizing fuel consumption and time of flight simultaneously. 
A trajectory is labeled Pareto-optimal if no other feasible solution attains strictly lower values for both objectives.
The propulsion system has maximum thrust $T_{max}$ and constant specific impulse $I_{sp}$ with exhaust velocity $c = I_{sp} g_0$, where $g_0$ is the standard gravitational acceleration.
The state vector $\boldsymbol{x}=(\boldsymbol{r}^\top,\boldsymbol{v}^\top, m) \in \mathbb{R}^{7}$ includes the spacecraft's position $\boldsymbol{r}\in\mathbb{R}^3$, velocity $\boldsymbol{v}\in\mathbb{R}^3$ and mass $m\in\mathbb{R}$.
It evolves according to the autonomous dynamical system:
\begin{equation}
\label{Equation: dynamics}
    \dot{\boldsymbol{x}} = \boldsymbol{f}(\boldsymbol{x},\boldsymbol{u}) =
	\begin{pmatrix}
	\dot{\boldsymbol{r}} \\
	\dot{\boldsymbol{v}} \\
	\dot{m}
	\end{pmatrix}
	=
	\begin{pmatrix}
	\boldsymbol{v} \\
	\boldsymbol{g}(\boldsymbol{r},\boldsymbol{v}) + \frac{T}{m}\hat{\boldsymbol{u}} \\
	-\frac{T}{c}
	\end{pmatrix},
\end{equation}
where the natural acceleration of the system is described by the vector field $\boldsymbol{g}$.
We consider the control vector of the system to consist of three components: $\boldsymbol{u}=(\hat{\boldsymbol{u}}^\top,T,\zeta)^\top$.
The thrust direction is described by the unit vector $\hat{\boldsymbol{u}}\in\mathbb{R}^3$ and its magnitude is controlled through the throttle $\sigma$, where $T=\sigma T_{max}$.
To avoid writing the bounds $\sigma \in [0,1]$ as inequality constraints, a slack variable $\zeta$ with $\sigma=\sin^2 \zeta$ is introduced.
This allows us to write the set of admissible controls as two equality path constraints:
\begin{equation}
    \boldsymbol{\xi}(\boldsymbol{x}(t), \boldsymbol{u}(t), t) = 
    \begin{pmatrix}
	\hat{\boldsymbol{u}}^\top\hat{\boldsymbol{u}} - 1 \\
    T-T_{max}\sin^2\zeta
    \end{pmatrix} = \boldsymbol{0}.
\end{equation}
The trajectory starts from a fixed initial state and targets a fixed terminal position and velocity:
\begin{equation}
\label{Equation: Boundary conditions trajectory}
	\boldsymbol{x}(t_0) = \boldsymbol{x}_0, \quad\boldsymbol{e}(\boldsymbol{x}(t_f)) =
    \begin{pmatrix}
	\boldsymbol{r}_f - \boldsymbol{r}(t_f) \\
    \boldsymbol{v}_f - \boldsymbol{v}(t_f) 
    \end{pmatrix} = \boldsymbol{0},
\end{equation}
where $\boldsymbol{e}$ denotes the constraint violation vector. 
The final time $t_f$ is a free variable.

\subsection{Primer Vector Theory}
Primer Vector Theory applies an indirect approach to the optimal control problem defined in Section \ref{sec: optimal control problem}, in the context of spacecraft trajectory optimization \cite{Rutherfobd.1964}.
Using techniques from calculus of variations, the first‑order necessary conditions for optimality are applied to reformulate Eq. \eqref{equation: cost functional} as a two-point boundary value problem \cite{Liberzon.2012}.  
The dimension of the state vector is doubled through the introduction of a costate vector $\boldsymbol{\lambda} =(\boldsymbol{\lambda}_r^T,\boldsymbol{\lambda}_v^T, \lambda_m)^T \in \mathbb{R}^{7}$, consisting of position, velocity and mass costates.
The Hamiltonian of the system is given by 
\begin{equation}
	\label{Equation: Problem Hamiltonian}
	H = \mathcal{L} + \boldsymbol{\lambda}^\top \boldsymbol{f}(\boldsymbol{x},\boldsymbol{u}) = \boldsymbol{\lambda}_r^\top \boldsymbol{v} + \boldsymbol{\lambda}_v^\top \left(\boldsymbol{g}(\boldsymbol{r},\boldsymbol{v}) + \frac{T}{m}\hat{\boldsymbol{u}}\right) - \lambda_m\frac{T}{c}.
\end{equation}
According to Pontryagin's Minimum principle \cite{Pontryagin.2018}, we set $\hat{\boldsymbol{u}} = -\boldsymbol{\lambda}_v/\lambda_v$, to minimize $H$.
Throughout this paper, writing a vector without boldface (e.g., $\lambda_v$) denotes its Euclidean norm.
By introducing a switching function $S = \lambda_v + \lambda_m m / c$ we reformulate the Hamiltonian as
\begin{equation}
	\label{Equation: Hamiltonian without control}
	H = \boldsymbol{\lambda}_r^\top \boldsymbol{v} + \boldsymbol{\lambda}_v^\top \boldsymbol{g}(\boldsymbol{r},\boldsymbol{v}) - S\frac{T}{m}.
\end{equation}
To minimize $H$, the throttle is chosen based on the sign of $S$, resulting in the discontinuous "bang-bang" control law \cite{Conway.2010}:
\begin{equation}
	\label{Equation: Control Law}
	\hat{\boldsymbol{u}} = -\frac{\boldsymbol{\lambda_v}}{\lambda_v}, \quad
	\sigma = 
	\begin{cases}
	0 & \text{if } S < 0 \\
	1 & \text{if } S > 0 \\
	0 \leq \sigma \leq 1 & \text{if } S = 0.
	\end{cases}
\end{equation}
Applying the first-order necessary conditions for optimality yields the costate dynamics \cite{Liberzon.2012}:
\begin{equation}
\label{Equation: Costate equation of motion}
\dot{\boldsymbol{\lambda}} = \begin{pmatrix}
\dot{\boldsymbol{\lambda}}_r \\
\dot{\boldsymbol{\lambda}}_v \\
\dot{\lambda}_m
\end{pmatrix} = \begin{pmatrix}
- \boldsymbol{G}^\top \boldsymbol{\lambda}_v \\
-\boldsymbol{\lambda}_r - \boldsymbol{H}^\top \boldsymbol{\lambda}_v \\
-\lambda_v T / m^2
\end{pmatrix}, \quad\text{where}\quad
\mathbf{G} = \frac{\partial \mathbf{g}}{\partial \mathbf{r}} \quad \text{and} \quad \mathbf{H} = \frac{\partial \mathbf{g}}{\partial \mathbf{v}}.
\end{equation}
The equations of motions for the combined state and costate vector $\boldsymbol{y}\in \mathbb{R}^{14}$ are therefore given by:
\begin{equation}
\label{Equation: Combined state equations}
    \dot{\boldsymbol{y}}=\boldsymbol{f}(\boldsymbol{y}) = 
    \begin{pmatrix}
    \dot{\boldsymbol{r}} \\
    \dot{\boldsymbol{v}} \\
    \dot{m} \\
    \dot{\boldsymbol{\lambda}}_r \\
    \dot{\boldsymbol{\lambda}}_v \\
    \dot{\lambda}_m
    \end{pmatrix} = \begin{pmatrix}
    \boldsymbol{v} \\
    \boldsymbol{g}(\boldsymbol{r},\boldsymbol{v}) - (\boldsymbol{\lambda}_v/\lambda_v) T/m \\
    -T/c \\
    - \boldsymbol{G}^T \boldsymbol{\lambda}_v \\
    -\boldsymbol{\lambda}_r - \boldsymbol{H}^T \boldsymbol{\lambda}_v \\
    -\lambda_v T / m^2
    \end{pmatrix}.
\end{equation}
Combined with \eqref{Equation: Boundary conditions trajectory} and Eq. \eqref{Equation: Control Law} this yields a two-point boundary value problem.
To fix the degrees of freedom introduced by the free final mass and final time, we set $\lambda_m(t_0)=-1$ and implicitly target the transversality condition
\begin{equation}
    \label{Equation: lam_m=-k}
     H(\boldsymbol{x}(t_f), \boldsymbol{u}(t_f), \boldsymbol{\lambda}(t_f), t_f)=0.
\end{equation}
The resulting problem can be solved numerically. 

\subsection{Analytic Derivatives}
\label{sec: AD}
Employing a gradient-based method to find feasible solutions of the two-point boundary-value problem requires the derivatives of the constraint violations and fuel consumption with respect to the costates at the initial time.
These derivatives are obtained by numerically propagating the state transition matrix (STM), defined as $\boldsymbol{\Phi}(t,t_0)=\pd{\boldsymbol{y}(t)}{\boldsymbol{y}_0}$.
The STM is governed by the matrix differential equation: 
\begin{equation}
\label{eq: MDE}
    \dot{\boldsymbol{\Phi}}(t,t_0)=\boldsymbol{A}(t)\boldsymbol{\Phi}(t,t_0), 
\end{equation}
with initial condition $\boldsymbol{\Phi}(t_0,t_0)=\boldsymbol{I}$.
We now introduce the Jacobian $\boldsymbol A(t)\in\mathbb{R}^{14\times 14}$ of the system in Eq. \eqref{Equation: Combined state equations} which is given by:

\begin{equation}
\label{eq: State Jacobian}
\boldsymbol A(t)=\frac{\partial \boldsymbol{f}}{\partial \boldsymbol{y}} =
\left(
\begin{array}{cccccc}
\boldsymbol{0} & \boldsymbol{I} & \boldsymbol{0} & \boldsymbol{0} & \boldsymbol{0} & \boldsymbol{0} \\[6pt]
\boldsymbol{G} & \boldsymbol{H} & \dfrac{\boldsymbol{\lambda}_v}{\lambda_v}\dfrac{T}{m^2} & \boldsymbol{0} & -\dfrac{T}{m}\!\left(\dfrac{\boldsymbol{I}}{\lambda_v} - \dfrac{\boldsymbol{\lambda}_v \boldsymbol{\lambda}_v^\top}{\lambda_v^3}\right)  & \boldsymbol{0} \\[6pt]
\boldsymbol{0} & \boldsymbol{0} & 0 & \boldsymbol{0} & \boldsymbol{0} & 0 \\[10pt]
-\dfrac{\partial (\boldsymbol{G}^\top \boldsymbol{\lambda}_v)}{\partial \boldsymbol{r}} & \boldsymbol{0} & \boldsymbol{0} & \boldsymbol{0} & -\boldsymbol{G}^\top & \boldsymbol{0} \\[10pt]
\boldsymbol{0} & -\dfrac{\partial (\boldsymbol{H}^\top \boldsymbol{\lambda}_v)}{\partial \boldsymbol{v}} & \boldsymbol{0} & -\boldsymbol{I} & -\boldsymbol{H}^\top & \boldsymbol{0} \\[10pt]
\boldsymbol{0} & \boldsymbol{0} & \dfrac{2 \lambda_v T}{m^3} &
\boldsymbol{0} &
-\dfrac{\boldsymbol{\lambda}_v^\top T}{\lambda_v m^2} & 0
\end{array},
\right).
\end{equation}
The Jacobian is valid for both thrust and coast arcs by setting $T=T_{max}$ and $T=0$ respectively. 
Note, however, that the STM mapping of small perturbations from the initial to final state is only valid on continuous trajectories. 
Switching from a thrust to a coast arc and vice versa introduces a total of $N$ discontinuities in between the $N+1$ arcs of the trajectory. 
To determine the sensitivity of the final state with respect to the initial state we apply the chain rule:
\begin{equation}
\frac{\partial \boldsymbol{y}(t_f)}{\partial \boldsymbol{y}(t_0)}
= \boldsymbol{\Phi}(t_f, t_{N+}) \boldsymbol{\Psi}_N \boldsymbol{\Phi}(t_{N-}, t_{(N-1)+}) 
\boldsymbol{\Psi}_{N-1} \cdots 
\boldsymbol{\Phi}(t_{2-}, t_{1+}) \boldsymbol{\Psi}_1 \boldsymbol{\Phi}(t_{1-}, t_0),
\end{equation}
where the mapping across a discontinuity at time $t_n$ is described by the matrix:
\begin{equation}
\label{eq: discontinuity}
\boldsymbol{\Psi}_n \equiv 
\frac{\partial \boldsymbol{y}(t_{n+})}{\partial \boldsymbol{y}(t_{n-})}
= \boldsymbol{I}_{14\times 14} 
+ (\dot{\boldsymbol{y}}|_{t_{n+}} - \dot{\boldsymbol{y}}|_{t_{n-}})
\bigl(\tfrac{\partial S}{\partial \boldsymbol{y}} / \dot{S}|_{t_{n-}}\bigr).
\end{equation}
Here we use $t_{n-}$ and $t_{n+}$ to describe the state immediately before and after the discontinuity. 
The additional non-identity term in Eq. \eqref{eq: discontinuity} originates from a change in the switching time due to a perturbation of the state at $t_{n-}$, which then results in a perturbation of the state at $t_{n+}$. 
A more detailed explanation is provided by Russell \cite{Russell.2007}.
Evaluating the derivatives yields:
\begin{equation}
\boldsymbol{\Psi}_n=\boldsymbol{I}_{14\times 14} + \frac{1}{\hat{\boldsymbol{\lambda}}_v^\top\dot{\boldsymbol{\lambda}}_v|_{t_n}+1/c[\dot{m}|_{t_{n-}}\lambda_m+\dot{\lambda}_m|_{t_{n-}}m]}
\begin{blockarray}{cccccc}
\begin{block}{[cccccc]}
\displaystyle
\boldsymbol{0} & \boldsymbol{0} & \boldsymbol{0} & \boldsymbol{0} & \boldsymbol{0} & \boldsymbol{0}\\[4pt]
\boldsymbol{0} &
\boldsymbol{0} &
\displaystyle \frac{\Delta T\,\lambda_m}{mc}\,\hat{\boldsymbol{\lambda}}_v &
\boldsymbol{0} &
\displaystyle \frac{\Delta T}{m}\,\hat{\boldsymbol{\lambda}}_v\,\hat{\boldsymbol{\lambda}}_v^{\!\top} &
\frac{\Delta T}{c}\,\hat{\boldsymbol{\lambda}}_v
\\[8pt]
\boldsymbol{0} &
\boldsymbol{0} &
\displaystyle \frac{\Delta T\,\lambda_m}{c^2} &
\boldsymbol{0} &
\displaystyle \frac{\Delta T}{c}\,\hat{\boldsymbol{\lambda}}_v^{\!\top} &
\displaystyle \frac{\Delta T\,m}{c^2}
\\[8pt]
\boldsymbol{0} & \boldsymbol{0} & \boldsymbol{0} & \boldsymbol{0} & \boldsymbol{0} & \boldsymbol{0}\\[4pt]
\boldsymbol{0} & \boldsymbol{0} & \boldsymbol{0} & \boldsymbol{0} & \boldsymbol{0} & \boldsymbol{0}\\[4pt]
\boldsymbol{0}&
\boldsymbol{0} &
\displaystyle \frac{\,\Delta T\,\lambda_m\,\lambda_v}{m^{2}c} &
\boldsymbol{0} &
\displaystyle \frac{\Delta T}{m^{2}}\,{\boldsymbol{\lambda}}_v^{\!\top} &
\displaystyle \frac{\Delta T\,\lambda_v}{mc}
\\
\end{block}
\end{blockarray},
\end{equation}
where $\hat{\boldsymbol{\lambda}}_v=\boldsymbol{\lambda}_v/\lambda_v$ and $m$, $\boldsymbol{\lambda}_v$ and $\lambda_m$ are all evaluated at time $t_n$ and
\begin{equation}
    \Delta T = T(t_{n-})-T(t_{n+})= \begin{cases}
        T_{max} \quad \text{when switching from thrust arc to coast arc} \\
        -T_{max} \quad \text{when switching from coast arc to thrust arc.}
    \end{cases}
\end{equation}
The required derivatives can then be extracted as the corresponding submatrices from the STM chain: 
\begin{equation}
\begin{aligned}
\boldsymbol{G}_1
&= \begin{bmatrix} 
\pd{\boldsymbol{e}}{\boldsymbol{\lambda}_r(t_0)} & \pd{\boldsymbol{e}}{\boldsymbol{\lambda}_v(t_0)} 
\end{bmatrix}=
- \begin{bmatrix}
\displaystyle \frac{\partial \boldsymbol{r}(t_f)}{\partial \boldsymbol{\lambda}_r(t_0)}
&
\displaystyle \frac{\partial \boldsymbol{r}(t_f)}{\partial \boldsymbol{\lambda}_v(t_0)}
\\
\displaystyle \frac{\partial \boldsymbol{v}(t_f)}{\partial \boldsymbol{\lambda}_r(t_0)}
&
\displaystyle \frac{\partial \boldsymbol{v}(t_f)}{\partial \boldsymbol{\lambda}_v(t_0)}
\end{bmatrix}
= 
-\left[
\frac{\partial \boldsymbol{y}(t_f)}{\partial \boldsymbol{y}(t_0)}
\right]_{1:6,\;8:13},
\\[14pt]
\boldsymbol{G}_2
&=
- \begin{bmatrix}
\displaystyle \frac{\partial m(t_f)}{\partial \boldsymbol{\lambda}_r(t_0)}
&
\displaystyle \frac{\partial m(t_f)}{\partial \boldsymbol{\lambda}_v(t_0)}
\end{bmatrix}
=
-\left[
\frac{\partial \boldsymbol{y}(t_f)}{\partial \boldsymbol{y}(t_0)}
\right]_{7,\;8:13}.
\end{aligned}
\end{equation}
We extract only the derivatives with respect to the position and velocity costates, since the initial mass costate is fixed.

The discontinuity mapping is often neglected, as it only contributes a small change in the STM if the trajectory does not include too many switches. 
However, for the multi-revolution transfers in this work, which often include more than $20$ switches, including this additional term is crucial for accurate derivatives. 
\subsection{Circular Restricted Three Body Problem}
The CR3BP is a widely used idealization for preliminary trajectory design in astrodynamics. 
In this model, two massive bodies govern the motion of a third body with negligible mass, so that only the gravitational influence of the larger bodies is considered. 
Let the two massive bodies have masses $m_1$ and $m_2$ with $m_1>m_2$, referred to as the primary and secondary, respectively.
We adopt the standard system of non-dimensional natural units (NU) in which distances, times, and masses are normalized: the distance unit (DU) equals the separation between the two primaries; the time unit (TU) is their orbital period divided by $2\pi$; and the mass unit (MU) is $m_1+m_2$. 
Under this normalization, the single dimensionless parameter of the problem is the mass ratio $\mu=m_2/(m_1+m_2)$.
The equations of motion are expressed in a uniformly rotating reference frame in which the primaries remain fixed on the $r_1$-axis at locations $r_1=-\mu$ and $r_1=1-\mu$ respectively. 
The resulting gravitational field is described by
\begin{equation}
    \label{Equation: CR3BP dynamical system}
    \ddot{\boldsymbol{r}} = 
    \begin{pmatrix}
        \ddot{r}_1 \\
        \ddot{r}_2 \\
        \ddot{r}_3 
    \end{pmatrix}
    = \boldsymbol{g}(\boldsymbol{r},\boldsymbol{v}) =
    \begin{pmatrix}
        2v_2 + r_1 - (1-\mu)\frac{r_1+\mu}{\rho_1^3} - \mu\frac{r_1-1+\mu}{\rho_2^3} \\
        -2v_1 + r_2 - (1-\mu)\frac{r_2}{\rho_1^3} - \mu\frac{r_2}{\rho_2^3} \\
        -(1-\mu)\frac{r_3}{\rho_1^3} - \mu\frac{r_3}{\rho_2^3}
    \end{pmatrix},
\end{equation}
where
$\rho_1= \sqrt{(r_1+\mu)^2+r_2^2+r_3^2}$ and $\rho_2=\sqrt{(r_1-1+\mu)^2+r_2^2+r_3^2}$ denote the distances to the primary and secondary, respectively.

\section{Methodology}
\subsection{Diffusion Models}
Diffusion models have recently emerged as a leading class of deep generative models. Originating from concepts in non-equilibrium thermodynamics, they were first introduced to machine learning by Sohl-Dickstein \cite{SohlDickstein.3122015} and subsequently advanced by Song and Ermon \cite{Song.7122019,DBLP:journals/corr/abs-2011-13456} and Ho et al.\cite{Ho.6192020}.
Since then, diffusion-based approaches have been successfully applied across a wide range of domains, including reinforcement learning \cite{Wang.8122022,Ding.252024}, motion generation \cite{Tevet.9292022} and trajectory optimization \cite{li2023amortizedglobalsearchtrajectory,Graebner.1032024}.

As part of the family of latent variable probabilistic models, diffusion models are capable of modeling complex, high-dimensional distributions.
They operate by gradually adding noise to data through a forward diffusion process and then training a neural network to learn the reverse process.
Below we give a brief overview of the basic ideas behind diffusion models. 
While the specific model employed in this work is based on the Denoising Diffusion Probabilistic Model (DDPM) from Ho et al. \cite{Ho.6192020}, we describe the underlying ideas from a score-based perspective \cite{DBLP:journals/corr/abs-2011-13456}. 
This allows us to write the forward and reverse process as stochastic differential equations (SDEs), which highlights the connection to the gradient-based MCMC algorithms described in Section \ref{sec: gradient-based MCMC}.
The discretized DDPM formulation can be interpreted as a fixed-step numerical integration of the continuous-time SDEs.

Given a dataset with a finite number of samples
$\mathcal{D}=\{\boldsymbol{z}_1(0), \dots, \boldsymbol{z}_N(0)\}$ with $\boldsymbol{z}_i(0)\sim p_0(\boldsymbol{z})$, the goal of generative modeling is to learn the underlying distribution $p_0(\boldsymbol{z})$ and subsequently generate new samples from it.
Score-based generative models achieve this by constructing a time-continuous diffusion process $\{\boldsymbol{z}(t)\}_{t=0}^T$, modeled by the SDE
\begin{equation}
    d\boldsymbol{z}_t=\boldsymbol{f}_{\text{drift}}(\boldsymbol{z}_t,t)dt+g(t)d\boldsymbol{w}_t,
\end{equation}
where $\boldsymbol{f}_{\text{drift}}(\boldsymbol{z}_t,t)$ is the drift coefficient, $g(t)$ the diffusion coefficient, and $\boldsymbol{w}_t$ standard Brownian motion.
For denoising diffusion models, $\boldsymbol{f}_{\text{drift}}$ and $g$ are chosen such that $\boldsymbol{z}(t) \sim \mathcal{N}(\boldsymbol{0},\boldsymbol{I})$ as $t\rightarrow\infty$.
Assuming $\boldsymbol{z}(T) \sim \mathcal{N}(\boldsymbol{0},\boldsymbol{I})$ for sufficiently large $T$, one can start with samples from a simple distribution and generate samples $\boldsymbol{z}(0)\sim p_0(\boldsymbol{z})$ by simulating the reverse-time SDE:
\begin{equation}
    d\boldsymbol{z}_t=-[\boldsymbol{f}(\boldsymbol{z},t)-g(t)^2 \nabla_{\boldsymbol{z}}\log p_t(\boldsymbol{z}_t)]dt+g(t)d\overline{\boldsymbol{w}}_t,
\end{equation}
where $\overline{\boldsymbol{w}}_t$ denotes reverse Brownian motion. 
The gradient $\nabla{\boldsymbol{z}}\log p_t(\boldsymbol{z}_t)$ is referred to as the \textit{score function}, with $p_t$ denoting the distribution of $\boldsymbol{z}_t$ at time $t$. 
The score function is unknown but can be estimated by $\boldsymbol{s}_{\theta}(\boldsymbol{z},t)$, which is learned by a neural network with parameters $\theta$.
The neural network is trained using denoising score matching with loss
\begin{equation} \label{eq: score based loss}
L^{\text{simple}}(\theta) 
= \mathbb{E}_{P} 
\left[ \left\| \boldsymbol{s}_{\theta}(\boldsymbol{z}(t),t) - \nabla_{\boldsymbol{z}}\log p_t(\boldsymbol{z}(t)|\boldsymbol{z}(0)) \right\|^2 \right],
\end{equation}
where $(\boldsymbol{z}(0), \boldsymbol{z}(t), t) \sim P^{\text{cont}}$ with the joint distribution $P^{\text{cont}} \equiv \mathcal{D}\times p_t(\cdot|\boldsymbol{z}(0))\times \text{Uniform}([0,T])$.

\begin{figure}[t!]
\centering
\includegraphics[width=.8\textwidth]{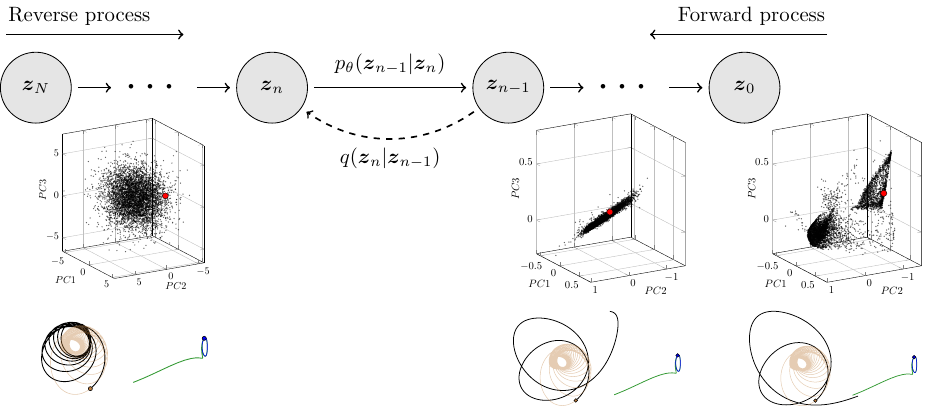}
\caption{Visualization of the forward and reverse diffusion processes for a spacecraft trajectory optimization problem. The distributions depict datasets of control vectors at various stages of the diffusion process with an example trajectory corresponding to the realization of the data-point marked in red.}
\label{fig: diffusion_process}
\end{figure}

The DDPM formulation employs an Euler-Maruyama discretization of the continuous-time Markov process by selecting $N$ integration points and introducing a cosine-based variance schedule $\{ \beta_n \in (0, 1) \}_{n=1}^N$.
The forward process is modeled by the Gaussian step
\begin{align} \label{eq: forward diffusion process step}
  q(\boldsymbol{z}_n | \boldsymbol{z}_{n-1}) = \mathcal{N}(\boldsymbol{z}_n; \sqrt{1 - \beta_n}\boldsymbol{z}_{n-1}, \beta_n \boldsymbol{I}),\quad
  q(\boldsymbol{z}_{n} | \boldsymbol{z}_{0}) = \mathcal{N}(\boldsymbol{z}_n; \sqrt{\overline{\alpha}_n}\boldsymbol{z}_{0}, (1-\overline{\alpha}_n)\boldsymbol{I}),
\end{align}
where $\boldsymbol{I}$ is the identity matrix, $\alpha_n = 1 - \beta_n$, and $\overline{\alpha}_n = \prod_{i=1}^n \alpha_i$. 
A simplified training loss is used:
\begin{equation} \label{eq: simplified loss}
L^{\text{simple}}(\theta) 
= \mathbb{E}_{P} 
\left[ \left\| \boldsymbol{\epsilon}_n - \boldsymbol{\epsilon}_{\theta}(\boldsymbol{z}_n, n) \right\|^2 \right],
\end{equation}
where $(\boldsymbol{z}_0, \boldsymbol{\epsilon}_n, n) \sim P \equiv \mathcal{D}\times\mathcal{N}(0, \mathbf{I})\times \text{Uniform}(\{1, \dots, N\})$. 
Here, $P$ denotes the joint distribution over the data sample $\boldsymbol{z}_0$ (drawn from the training dataset $\mathcal{D}$), the Gaussian noise 
$\boldsymbol{\epsilon}_n=\mathcal{N}(\boldsymbol{0},\boldsymbol{I})$, and uniform samples of the discrete time index $n$.
During sampling, the reverse-time SDE is solved, with one step given by 
\begin{equation} \label{eq: sampling}
	\boldsymbol{z}_{n-1} = \frac{1}{\sqrt{\alpha_n}} \left( \boldsymbol{z}_n - \frac{1 - \alpha_n}{\sqrt{1 - \alpha_n}} \boldsymbol{\epsilon}_{\theta}(\boldsymbol{z}_n,n) \right) + \sqrt{\tilde{\beta}_ n}\boldsymbol{y}(n) \quad
	\text{ where } \quad \boldsymbol{y}(n) 
	\begin{cases}
		=\boldsymbol{0} & \text{for } n = 1 \\
		\sim \mathcal{N}(\boldsymbol{0}, \boldsymbol{I}) & \text{otherwise,}
	\end{cases}
\end{equation}
and $\tilde{\beta}_n = (1 - \overline{\alpha}_{n-1})/(1 - \overline{\alpha}_n)\cdot \beta_n $.
This step can also be expressed as the transition probability $p_{\theta}(\boldsymbol{z}_{n-1}|\boldsymbol{z}_n)$.
Figure \ref{fig: diffusion_process} visualizes the forward and reverse processes for a spacecraft trajectory optimization example, where the data distribution is defined over the costate space of high-quality solutions.

\subsection{Supervised Fine-tuning}
For both large language models and text-to-image systems, prior work has shown that fine-tuning with human feedback can significantly improve generative performance \cite{Ouyang.342022,Lee.2232023,Ziegler.9182019}.
In these methods, a separate reward model is first trained from human evaluations of generated samples. 
This reward model is then used to score outputs from a pre-trained baseline model, and those scores guide the fine-tuning procedure through reward-weighted likelihood maximization \cite{Lee.2232023}.

In our setting, we adopt a similar strategy for fine-tuning diffusion models that generate initial costates for trajectory optimization.
However, unlike the human-feedback paradigm, the reward function $R(\boldsymbol{z})$ is directly computable, as detailed in Section \ref{sec: self-supervided training framework}, eliminating the need to train a separate reward model.
The model parameters are initialized from the baseline model and updated by minimizing the loss function \cite{Lee.2232023}
\begin{equation}
L(\theta) = \mathbb{E}_{P^{\text{new}}}  \Big[ R(\boldsymbol{z}_0)\,\Big\| \boldsymbol{\epsilon}_n - \boldsymbol{\epsilon}_{\theta}(\boldsymbol{z}_n, n) \Big\|^2 \Big]
\label{eq:updated loss}
\end{equation}
where $P^{\text{new}}$ is a product distribution in the same sense as $P$ and samples from the new training data $\mathcal{D}^{\text{new}}$.

\subsection{Preliminary Screening Algorithm}
We employ a modified version of a preliminary screening algorithm described in the author's previous work \cite{graebner_JAS} to evaluate the objective value of a given costate sample.
In the remainder of this work, $\boldsymbol{\lambda}$ refers specifically to the costate vector at initial time, with the mass costate excluded.
Starting from the fixed initial state and a given initial costate, the trajectory is  propagated according to the system dynamics in Eq. \eqref{Equation: Combined state equations} for a shooting time $\tau_{s,\text{max}}$. 
Numerical integration is performed using \texttt{pydylan}, the Python interface to the astrodynamics software package Dynamically Leveraged (N) Multibody Trajectory Optimization (DyLAN) \cite{Beeson.Aug.2022}, which employs an adaptive-stepsize RK54 integrator.
The trajectory is parameterized by the shooting time $\tau_s$, while the target point on the final orbit is parameterized by a coast time $\tau_f$.
This coast time $\tau_f$ corresponds to the time of flight required to reach a particular point on the target orbit and is limited by the orbit period $\mathcal{T}_f$. 
It enables us to transcribe position and velocity on the orbit as $\boldsymbol{r}_f(\tau_f)$ and $\boldsymbol{v}_f(\tau_f)$.
Each pair consisting of a trajectory state at time $\tau_s$ and an orbit state at time $\tau_f$ is assigned an objective value:
\begin{equation}
    \label{eq: general objective}
    J(\boldsymbol{\lambda},\tau_s,\tau_f)=e(\boldsymbol{\lambda},\tau_s,\tau_f)+\kappa_1 \left(\frac{\Delta m(\boldsymbol{\lambda},\tau_s)}{m_0}+\kappa_2\tau_s\right)
\end{equation}
with 
\begin{equation}
\label{eq: constr violation}
e(\boldsymbol{\lambda},\tau_s,\tau_f)=\left\Vert
    \begin{pmatrix}
     \boldsymbol{r}_f(\tau_f) - \boldsymbol{r}(\tau_s) \\
    \boldsymbol{v}_f(\tau_f) -\boldsymbol{v}(\tau_s)
    \end{pmatrix}\right\Vert_2.
\end{equation}
This objective function targets feasibility through the $\ell_2$ norm of the constraint violations $e$, while also including the hybrid objective of fuel consumption $\Delta m=m_0-m_f(\boldsymbol{\lambda},\tau_s)$
and time of flight. 
The trade-off between feasibility and optimality is controlled by the scaling parameter $\kappa_1$, while the two terms in the hybrid objective are weighted using a second scaling factor $\kappa_2$.

For a given sample $\boldsymbol{\lambda}$, the algorithm selects optimal shooting and coast times
\begin{equation}
    (\tau_s^*(\boldsymbol{\lambda}),\,\tau_f^*(\boldsymbol{\lambda}))
    = \argmin_{\substack{0 \le \tau_s \le \tau_{s,\max} \\[2pt] 0 \le \tau_f \le \mathcal{T}_f}}
      J(\boldsymbol{\lambda},\, \tau_s,\, \tau_f),
\end{equation}
which, when substituted into the Eq. \eqref{eq: general objective}, yields the reduced objective function:
\begin{equation}
    \label{eq:objective function}    J^*(\boldsymbol{\lambda})=J(\boldsymbol{\lambda},\tau_s^*(\boldsymbol{\lambda}),\tau_f^*(\boldsymbol{\lambda}))=\min\limits_{\substack{0\leq\tau_s\leq\tau_{s,\text{max}} \\ 0\leq\tau_f\leq\mathcal{T}_f}} J(\boldsymbol{\lambda},\tau_s,\tau_f).
\end{equation}
Within the algorithm, this minimization is carried out as a nearest-neighbor search over discretized sets of states on the trajectory and final orbit. 
It is executed efficiently using a k-dimensional (k-d) tree search \cite{Bentley.1975}.
Our goal is to find samples that minimize the reduced objective function $J^*$, which is henceforth referred to as the objective. 

\subsection{Gradient Approximation}
\label{sec: gradient approximation}
\begin{figure}[t!]
\centering\includegraphics[width=0.8\textwidth]{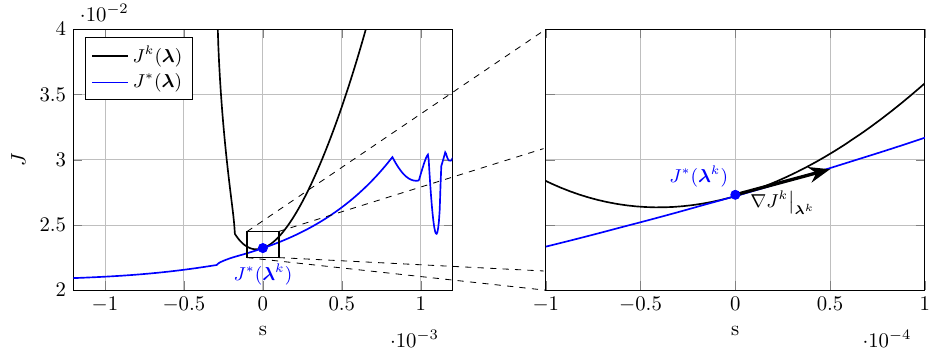}
	\caption{$J^k$ and $J^*$ for a randomly selected sample $\boldsymbol{\lambda}^k$. The objective values are plotted along the line defined by $\boldsymbol{\lambda}^k+s\nabla J^k(\boldsymbol{\lambda})|_{\boldsymbol{\lambda}=\boldsymbol{\lambda}^k}$, where $s\in\mathbb{R}$. The plot on the right is a close-up of the left plot around $s=0$.}
	\label{fig:Grad_approx}
\end{figure}
A gradient-based MCMC algorithm requires the gradient $\nabla J^*(\boldsymbol{\lambda})$ evaluated at the current costate $\boldsymbol{\lambda}^k$ at iteration $k$.
While the function $J^*$ is continuous, it is not continuously differentiable, and the gradient may not exist everywhere. 
By fixing the shooting and coast times at the given sample, $\tau_s^k=\tau_s^*(\boldsymbol{\lambda}^k)$ and $\tau_f^k=\tau_f^*(\boldsymbol{\lambda}^k)$, we define the frozen-time objective
\begin{equation}
J^k(\boldsymbol{\lambda})=J(\boldsymbol{\lambda},\tau_s^k,\tau_f^k)
\end{equation}
which is continuously differentiable in $\boldsymbol{\lambda}$.

In the following, we show that $\boldsymbol{d}=-\nabla J^k(\boldsymbol{\lambda})|_{\boldsymbol{\lambda}=\boldsymbol{\lambda}^k}$ is a descent direction for $J^*$ at $\boldsymbol{\lambda}^k$ if $\boldsymbol{d}\neq 0$.
First, note that by definition
\begin{equation}
    \label{eq: step 1}
    J^*(\boldsymbol{\lambda}^k)=J^k(\boldsymbol{\lambda}^k),
\end{equation}
and
\begin{equation}
    \label{eq: step 2}J^*(\boldsymbol{\lambda})\leq J^k(\boldsymbol{\lambda}) \quad \forall \boldsymbol{\lambda}.
\end{equation}
By definition of the gradient, $\boldsymbol{d}$ is a descent direction of $J^k$ at $\boldsymbol{\lambda}^k$, which means 
\begin{equation}
\label{eq: step 3}
\exists\overline{\alpha} \quad \text{such that} \quad
    J^k(\boldsymbol{\lambda}^k+\alpha \boldsymbol{d})<J^k(\boldsymbol{\lambda}^k) \quad \forall\alpha\in(0,\overline{\alpha}).
\end{equation}
Based on Eqs. \eqref{eq: step 2}, \eqref{eq: step 3} and \eqref{eq: step 1}, we obtain
\begin{equation}
    J^*(\boldsymbol{\lambda}^k+\alpha \boldsymbol{d})\leq J^k(\boldsymbol{\lambda}^k+\alpha \boldsymbol{d})<J^k(\boldsymbol{\lambda}^k)=J^*(\boldsymbol{\lambda}^k)\quad \forall\alpha\in(0,\overline{\alpha}),
\end{equation}
which establishes $\boldsymbol{d}$ as a descent direction of $J^*$.
Figure \ref{fig:Grad_approx} shows $J^*$ and $J^k$ for a random sample $\boldsymbol{\lambda}^k$, plotted along a line in the objective space. 
The zoomed view on the right clearly shows that $\nabla J^k(\boldsymbol{\lambda})|_{\boldsymbol{\lambda}=\boldsymbol{\lambda}^k}$ is a good approximation of $\nabla J^*(\boldsymbol{\lambda})|_{\boldsymbol{\lambda}=\boldsymbol{\lambda}^k}$ for this particular sample.

The gradient of the frozen-time objective $\nabla J^k(\boldsymbol{\lambda})|_{\boldsymbol{\lambda}=\boldsymbol{\lambda}^k}$ can be directly expressed in terms of the submatrices of the STM derived in Section \ref{sec: AD}:
\begin{align}
    \nabla J^k(\boldsymbol{\lambda})|_{\boldsymbol{\lambda}=\boldsymbol{\lambda}^k}&=\nabla e(\boldsymbol{\lambda},\tau_s^k,\tau_f^k)|_{\boldsymbol{\lambda}=\boldsymbol{\lambda}^k}-\dfrac{\kappa_1}{m_0}\nabla m_f(\boldsymbol{\lambda},\tau_s ^k)|_{\boldsymbol{\lambda}=\boldsymbol{\lambda}^k} \\
    &=\dfrac{1}{e(\boldsymbol{\lambda}^k,\tau_s^k,\tau_f^k)}\boldsymbol{G}_2^{\!\top}\boldsymbol{e}(\boldsymbol{\lambda}^k,\tau_s^k,\tau_f^k)+\dfrac{\kappa_1}{m_0}{\boldsymbol{G}_1}^{\!\top}.
\end{align}
Employing analytic derivatives derived from the STM achieves improved accuracy over a numerical finite differencing approach, which is particularly important for multi-revolution transfers that are highly sensitive to small perturbations \cite{Russell.2007}. 
For a problem with $d$ states, the STM computation increases the dimension of propagated states by $d^2$.

\subsection{Gradient-based MCMC}
\label{sec: gradient-based MCMC}
The novel contribution of this work is the inclusion of gradient information of the target distribution in the MCMC algorithm. 
In general, MCMC methods generate samples from a target distribution with an unnormalized density $\pi(\boldsymbol{\lambda})$ by constructing a Markov chain whose stationary distribution coincides with $\pi(\boldsymbol{\lambda})$.
Standard MCMC methods such as the random-walk Metropolis algorithm explore the search space slowly and struggle with multimodal distributions where the modes are well separated \cite{Roberts.1998, Tran.4202025}. 
Gradient-based samplers account for local geometry by including $\nabla\log \pi(\boldsymbol{\lambda})$ in the proposal distribution. 

We consider two variants of gradient-based MCMC, which are discussed in the following sections.
Both follow the general idea of MCMC algorithms by proposing a new state from a distribution $q$ that is easy to sample from and then accepting or rejecting that proposal based on an acceptance probability $\alpha$.

\subsubsection{Metropolis-Adjusted Langevin Algorithm}
The Metropolis-Adjusted Langevin Algorithm (MALA), also referred to as Langevin Monte Carlo, proposes new states using Langevin dynamics and accepts them based on the Metropolis-Hastings algorithm \cite{Roberts.1998, Metropolis.1953, Hastings.1970}.
Langevin dynamics are described by the stochastic differential equation
\begin{equation}
d\boldsymbol{\lambda}_t
= \nabla \log \pi(\boldsymbol{\lambda}_t)\, dt
\;+\; \sqrt{2}\, d\boldsymbol{w}_t ,
\end{equation}
where $\boldsymbol{w}_t$ is standard Brownian motion.
As $t\rightarrow\infty$, the probability distribution of $\boldsymbol{\lambda}(t)$ converges to the stationarity distribution $\pi$, which is invariant under the diffusion \cite{Roberts.1996}.
A first-order Euler-Maruyama discretization  of the SDE yields the standard MALA proposal distribution
\begin{equation}
    \label{eq: proposal distribution basic}
    q(\tilde{\boldsymbol{\lambda}}^{k+1}; \boldsymbol{\lambda}^k) 
    \equiv \mathcal{N}(\tilde{\boldsymbol{\lambda}}^{k+1}; \boldsymbol{\lambda}^k+\dfrac{\epsilon}2\nabla\!\log(\pi(\boldsymbol{\lambda}^{k})),\epsilon\boldsymbol{\text{I}}).
\end{equation}
with timestep $\epsilon$ and identity matrix $\boldsymbol{I}$.
This is basic MALA proposal performs poorly if the parameter space exhibits anisotropic scaling.
We therefore use the modified proposal 
\begin{equation}
    \label{eq: proposal distribution}
    q(\tilde{\boldsymbol{\lambda}}^{k+1}; \boldsymbol{\lambda}^k)
    \equiv 
    \mathcal{N}\!\left(
        \tilde{\boldsymbol{\lambda}}^{k+1};
        \boldsymbol{\lambda}^k
        + \frac{\epsilon}{2}\,\boldsymbol{\Sigma}_{\lambda}^{1/2}
        \frac{\nabla \log\!\pi(\boldsymbol{\lambda}^k)}
        {\left\lVert \nabla \log\!\pi(\boldsymbol{\lambda}^k) \right\rVert_2},
        \;\boldsymbol{\Sigma}_{\lambda}
    \right).
\end{equation}
where the proposal covariance $\boldsymbol{\Sigma}_{\lambda}$ is ideally approximately proportional to the covariance of the target distribution. 
Based on empirical results, we normalize the gradient to prevent excessively large drift steps in regions where the objective landscape is very steep.
For $\epsilon=0$ this proposal reduces exactly to the random-walk Metropolis algorithm.
For $\epsilon \neq0$ the method augments the random walk with a gradient drift step before drawing the stochastic proposal.  
The modified proposal allows us to control the size of the gradient step and the random step independently, which is not possible for the standard proposal in Eq. \eqref{eq: proposal distribution basic}.
A proposed state is accepted with probability
\begin{equation}
    \alpha(\boldsymbol{\lambda}^k,\tilde{\boldsymbol{\lambda}}^{k+1})=\frac{g(\tilde{\boldsymbol{\lambda}}^{k+1})q\bigl(\boldsymbol{\lambda}^{k};\tilde{\boldsymbol{\lambda}}^{k+1}\bigr)}{g(\boldsymbol{\lambda}^{k})q\bigl(\tilde{\boldsymbol{\lambda}}^{k+1};\boldsymbol{\lambda}^{k}\bigr)}\wedge1.
\end{equation}

\subsubsection{Hamiltonian Monte Carlo}
Hamiltonian Monte Carlo (HMC) is another gradient-based method that uses Hamiltonian dynamics to construct an MCMC algorithm. 
By computing new states along trajectories under these dynamics, correlation between successive samples is reduced and exploration of the state space is accelerated.
Because the trajectories follow regions of high target density, proposed states are accepted with high probability \cite{Neal.2012}.
We introduce the potential energy of the target distribution as $U(\boldsymbol{\lambda})=-\log \pi(\boldsymbol{\lambda})$.
By augmenting the state with additional momentum variables $\boldsymbol{p}\in\mathbb{R}^d$, where $d$ is the dimension of $\boldsymbol{\lambda}$, and introducing a kinetic energy term $K(\boldsymbol{p})$, the Hamiltonian of the system is defined as 
\begin{equation}
    H(\boldsymbol{\lambda},\boldsymbol{p})=U(\boldsymbol{\lambda})+K(\boldsymbol{p}).
\end{equation}
Common practice is to choose a homogeneous quadratic kinetic energy $K(\boldsymbol{p})=\frac{1}{2}\boldsymbol{p}^\top \boldsymbol{M}^{-1} \boldsymbol{p}$, with the mass matrix $M$ typically taken to be diagonal \cite{Neal.2012}.
The specific choice of the diagonal elements $\boldsymbol{M}_{ii}$ is one of the substantial degrees of freedom in HMC.
In practice, a good approach is to set $\boldsymbol{M}^{-1}$ proportional to the covariance of the target distribution or to an estimate thereof \cite{betancourt2018conceptualintroductionhamiltonianmonte}. 
The dynamics of the augmented state are described by Hamilton's equations:
\begin{equation}
\frac{\mathrm{d}\boldsymbol{\lambda}}{\mathrm{d}t}
  = \frac{\partial H}{\partial \boldsymbol{p}}
  = \nabla_{\boldsymbol{p}} K=\boldsymbol{M}^{-1} \boldsymbol{p},
\qquad
\frac{\mathrm{d}\boldsymbol{p}}{\mathrm{d}t}
  = -\,\frac{\partial H}{\partial \boldsymbol{\lambda}}
  = \nabla\log \pi(\boldsymbol{\lambda})
\end{equation}

The first step of HMC is to sample the initial momentum at the current state from the normal distribution $\boldsymbol{p}^k\sim\mathcal{N}(\boldsymbol{p};\boldsymbol{0},\boldsymbol{M})$.
In the second step, a new state $(\tilde{\boldsymbol{\lambda}}^{k+1},\tilde{\boldsymbol{p}}^{k+1})$ is proposed by integrating the current state $(\boldsymbol{\lambda}^{k},\boldsymbol{p}^{k})$ under the dynamics defined by the Hamiltonian.
Numerically, this is carried out using a volume-preserving symplectic integrator such as the Leapfrog method \cite{Neal.2012}. 
As in the MALA method, this step requires the gradient $\nabla \log\pi(\boldsymbol{\lambda})$.
One iteration of the leapfrog method consists of the following steps:
\begin{align}
\boldsymbol{p}^k\!\left(t + \frac{\Delta t}{2}\right)
&= \boldsymbol{p}^k(t)
  + \frac{\Delta t}{2}\,\nabla \log\pi(\boldsymbol{\lambda})
    \big|_{\boldsymbol{\lambda}=\boldsymbol{\lambda}^k(t)}
\label{eq:hmc_step1} \\
\boldsymbol{\lambda}^k(t+\Delta t)
&= \boldsymbol{\lambda}^k(t)
  + \Delta t\, \boldsymbol{M}^{-1}\,
    \boldsymbol{p}^k\!\left(t + \frac{\Delta t}{2}\right)
\label{eq:hmc_step2} \\
\boldsymbol{p}^k(t+\Delta t)
&= \boldsymbol{p}^k\!\left(t + \frac{\Delta t}{2}\right)
  + \frac{\Delta t}{2}\,\nabla \log\pi(\boldsymbol{\lambda})
    \big|_{\boldsymbol{\lambda}=\boldsymbol{\lambda}^k(t+\Delta t)}
\label{eq:hmc_step3}
\end{align}
with initial samples $\boldsymbol{p}^k(0)=\boldsymbol{p}^k$ and $\boldsymbol{\lambda}^k(0)=\boldsymbol{\lambda}^k$.
The integration runs for a total time of $L\Delta t$, where $L$ is the number of integration steps and $\Delta t$ is the integration timestep. 
The proposed state for the Metropolis-Hastings step is given by $\tilde{\boldsymbol{\lambda}}^{k+1}=\boldsymbol{\lambda}^{k}(L\Delta t)$ and $\tilde{\boldsymbol{p}}^{k+1}=\boldsymbol{p}^{k}(L\Delta t)$.
The acceptance probability of this new state is then
\begin{equation}
    \alpha(\boldsymbol{\lambda}^k,\tilde{\boldsymbol{\lambda}}^{k+1})=\frac{\exp{(-H(\tilde{\boldsymbol{\lambda}}^{k+1},\tilde{\boldsymbol{p}}^{k+1}))}}{\exp(-{H(\boldsymbol{\lambda}^{k},\boldsymbol{p}^{k}))}}\wedge1.
\end{equation}
Based on empirical observations, we adapt the standard HMC update to use the same normalized gradient scaling employed in the MALA method. 
Specifically, in Eqs. \ref{eq:hmc_step1} and \ref{eq:hmc_step3}, we replace $\nabla \log\pi(\boldsymbol{\lambda})$ by $\sqrt{\boldsymbol{M}}\nabla \log\pi(\boldsymbol{\lambda})/\left\lVert \nabla \log\!\pi(\boldsymbol{\lambda}^k) \right\rVert_2$.
With this modification, the HMC proposal becomes exactly equivalent to a MALA proposal when $L=1$, $\Delta t=\epsilon$ and $\boldsymbol{M}=\epsilon^2\boldsymbol{\Sigma}_\lambda^{-1}$. 
To keep the methods comparable, we parameterize the HMC algorithm direclty in terms of $\epsilon$ and $\boldsymbol{\Sigma}_{\lambda}$, and then compute $\Delta t$ and $\boldsymbol{M}$ from these quantities.

\subsection{Self-supervised Training Framework}
\label{sec: self-supervided training framework}
\begin{figure}[b!]
\centering\includegraphics[width=0.8\textwidth]{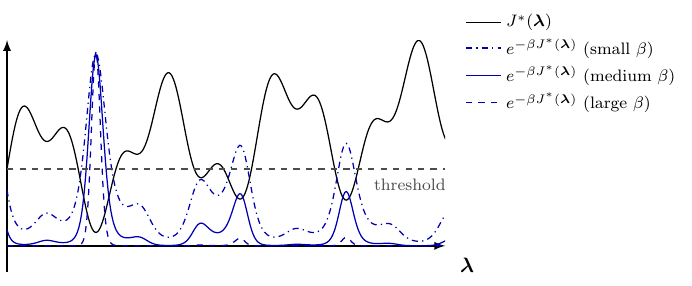}
	\caption{Reformulating optimization into sampling: minima of $J^*(\boldsymbol{\lambda})$ correspond to peaks of the unnormalized target density $\pi_{\alpha}(\boldsymbol{\lambda})\equiv\exp(-\beta J^*(\boldsymbol{\lambda}))$. The scaling factor $\beta$ controls the thickness and magnitude of peaks.}
	\label{fig: sampling function}
\end{figure}

The overall structure of the self-supervised diffusion model training framework follows our previous work, and we refer the reader to \cite{graebner2025aas} for a comprehensive description.
Here, we provide a brief overview and highlight the novel components introduced in this study.
The general idea of the framework is illustrated in Figure \ref{fig: MCMC_framework}.
The target distribution 
\begin{equation}
    \pi_{\alpha}(\boldsymbol{\lambda})\equiv\exp(-\beta J^*(\boldsymbol{\lambda})),
\end{equation}
is constructed to generate high-quality initial costates for an indirect spacecraft trajectory optimization problem with parameters $\alpha$.
A scaling factor $\beta$ controls the thickness of the high-density regions and the steepness of the gradient, as visualized in Figure \ref{fig: sampling function}.

We start from a baseline diffusion model, which generates samples from a distribution $p_{\tilde{\alpha}}$.
This model was trained on a transfer with different parameters $\tilde{\alpha}$, but ideally exhibits similar characteristics to our target distribution.
It serves as the initial distribution for the gradient-based MCMC step, which is detailed in section \ref{sec: gradient-based MCMC}.
We run $N$ Markov chains simultaneously to achieve both local and global exploration of the solution space. 
The drift term in the Markov proposal is approximated as described in Section \ref{sec: gradient approximation}:
\begin{equation}
    \nabla\log \pi(\boldsymbol{\lambda})=-\beta \nabla J^*(\boldsymbol{\lambda})\approx -\beta\nabla J^k(\boldsymbol{\lambda}),
\end{equation}
which allows us to compute the derivatives analytically from the STM.
\begin{figure}[t!]
\centering\includegraphics[width=0.95\textwidth]{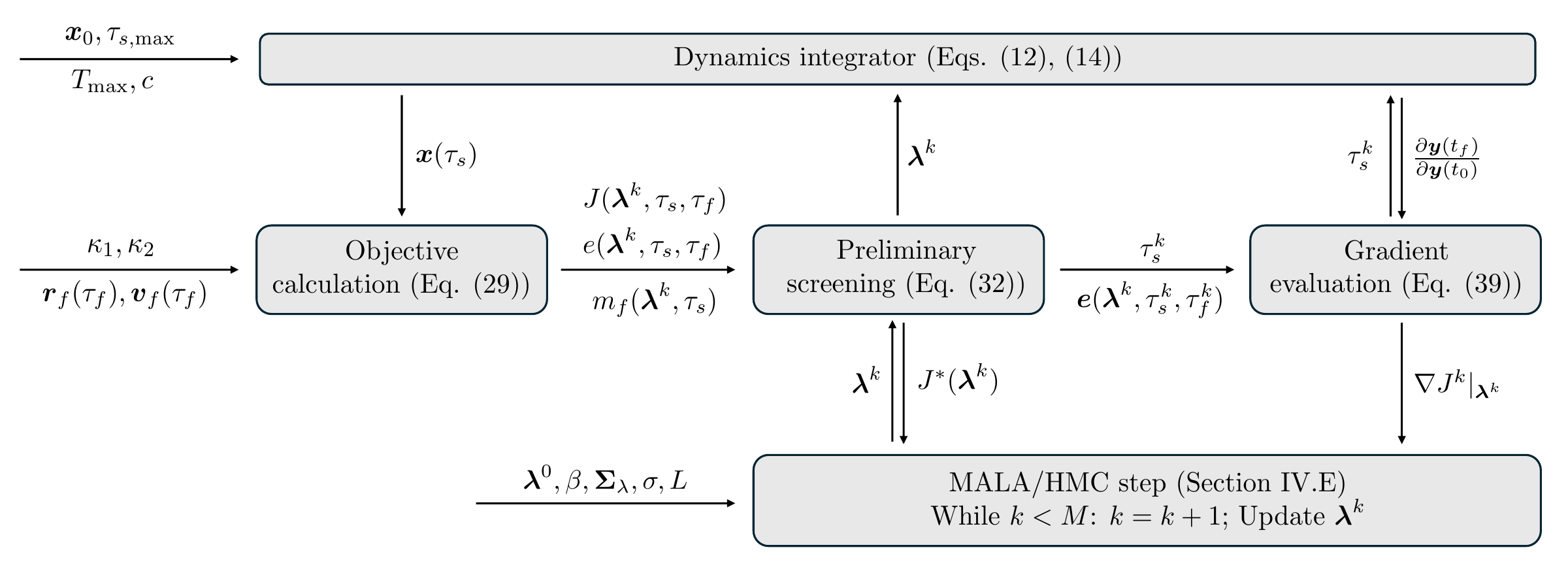}
	\caption{Flowchart of variables and equations for gradient‐based MCMC sampling.}
	\label{fig: Mala workflow}
\end{figure}
Each chain runs for a total of $M$ iterations, and the first $M_0$ burn-in samples are discarded.
Given a sufficiently long runtime and number of samples, the Markov chain converges to the target distribution and therefore consists of high-quality samples for the initial costates. 
We then evaluate the reward function
\begin{equation}
    \label{eq: reward}
    R(\boldsymbol{\lambda})\equiv a\exp(-b J^*(\boldsymbol{\lambda})),
\end{equation}
for each sample.
Similarly to the target distribution, the reward depends on the objective, with problem specific parameters $a$ and $b$.
These parameters are chosen such that $R(\boldsymbol{\lambda})$ spans the range $[0.1,1]$; thereby assigning the best sample ten times the weight of the worst.
The costate-reward pairs are then used to fine-tune the diffusion model using reward-weighted likelihood maximization.
The refined model is able to generate outputs from a distribution $p_{\alpha}$ that exhibits an improved fit to the target distribution.

\section{Results}
\begin{table}[b!]
\centering
\caption{Problem parameters for Europa and Titan DRO transfers.}
\setlength{\tabcolsep}{6pt}
\begin{tabular}{lcc}
\toprule
\textbf{Trajectory parameters} & \textbf{Europa DRO} & \textbf{Titan DRO} \\
\midrule
Initial state $[\boldsymbol{r}_0^T,\boldsymbol{v}_0^T]$ [NU] 
    & $[1.0752, 0.0, 0.0, 0.0, -0.1499, 0.0]$
    & $[1.0758, 0.0, 0.0, 0.0, -0.1684, 0.0]$ \\
Terminal state $[\boldsymbol{r}_f^T,\boldsymbol{v}_f^T]$ [NU] 
    & $[1.0306, 0.0, 0.0, 0.0, -0.0727, 0.0]$
    & $[1.0304, 0.0, 0.0, 0.0, -0.1248, 0.0]$ \\
Orbital period target DRO $\mathcal{T}_f$ [TU] 
    & 4.1055 & 4.6558 \\
Max.\ shooting time $\tau_{s,\max}$ [TU] 
    & \multicolumn{2}{c}{90} \\
\midrule
\textbf{Spacecraft parameters} &  &  \\
\midrule
Initial mass $m_0$ [kg] 
    & \multicolumn{2}{c}{25,000} \\
Fuel mass [kg] 
    & \multicolumn{2}{c}{15,000} \\
Dry mass [kg] 
    & \multicolumn{2}{c}{10,000} \\
Specific impulse $I_{sp}$ [s] 
    & 7,365 & 2,987 \\
Thrust magnitude $T_{max}$ [N] 
    & 4.735 & 0.4500 \\
\midrule
\textbf{Natural units} & \textbf{Jupiter–Europa} & \textbf{Saturn–Titan} \\
\midrule
Distance unit [km] 
    & 670,900 & 1,221,870 \\
Time unit [s] 
    & 48,822.76 & 219,277.51 \\
Mass unit [kg] 
    & $1.898 \times 10^{27}$ & $5.685 \times 10^{26}$ \\
Mass parameter $\mu$ 
    & $2.528 \times 10^{-5}$ & $2.366 \times 10^{-4}$ \\
\bottomrule
\end{tabular}
\label{tab:comparison_Europa_Titan}
\end{table}
Results are presented for a low-thrust transfer in the CR3BP. 
We benchmark the two gradient‑based samplers against each other and against the original random-walk Metropolis approach. 
Performance is evaluated in terms of feasibility ratio, objective quality, sample diversity and computational cost. 

For numerical propagation of trajectories and STMs, we employ an adaptive-stepsize RK54 integrator with relative tolerance $10^{-12}$ and maximum stepsize $10^{-2}$ in NU.
The switching function is interpolated to an accuracy of $10^{-13}$. 

\subsection{Problem Description}
\begin{figure}[t!]
  \centering
  %
  \begin{subfigure}[b]{0.45\textwidth}  
    \centering
    \includegraphics[height=5.9cm]{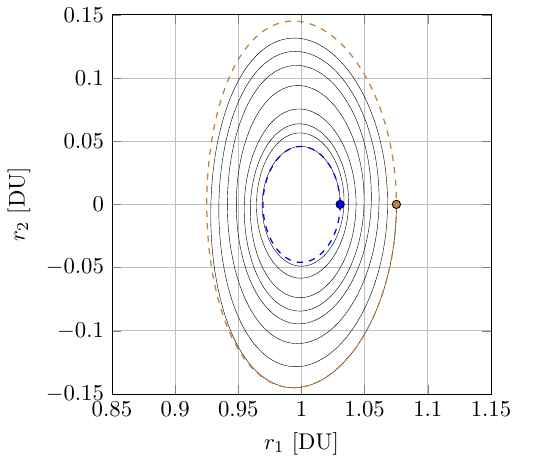}
  \end{subfigure}
  \hfill
  \begin{subfigure}[b]{0.54\textwidth}  
    \centering
    \includegraphics[height=5.9cm]{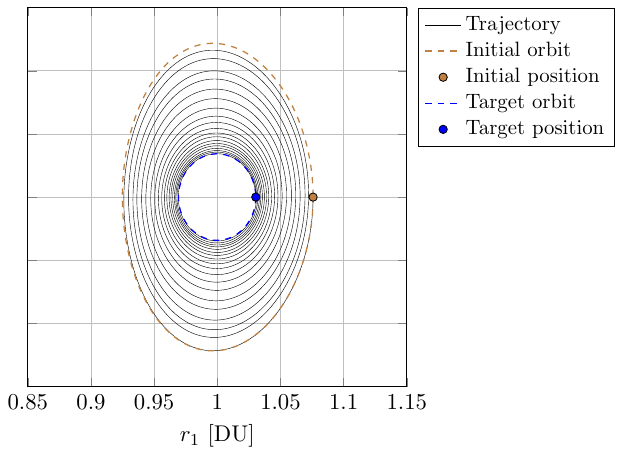}
  \end{subfigure}

  \caption{Example trajectories for the Europa DRO transfer (left) and Titan DRO transfer (right).}
  \label{fig:traj_examples}
\end{figure}

We test the framework on one of the problems from our previous work \cite{graebner2025aas}.
Starting with samples from a baseline diffusion model trained on a transfer in the Jupiter-Europa system, the MCMC algorithms target a related transfer in the Saturn-Titan system.
The solutions to both problems are planar multi-revolution trajectories from a high distant retrograde orbit (DRO) to a lower DRO around the secondary body. 
All relevant problem parameters, as well as the natural units for both systems, are presented in Table \ref{tab:comparison_Europa_Titan}.
The spacecraft parameters are chosen to be identical in natural units, leading to different values when converted to SI units. 
Initial and target DROs are selected to have the same distance to the secondary body in NU at the $r_1$-axis crossing closer to the primary in both systems.
The corresponding $v_2$ velocities are determined through a differential correction algorithm to achieve a closed orbit.
With the mass parameter $\mu$ in the Saturn-Titan system being an order of magnitude larger, the resulting DROs have qualitatively different shapes, as shown in Figure \ref{fig:traj_examples}.
The example trajectories in Figure \ref{fig:traj_examples} exhibit markedly different characteristics, with an average of $23.7$ revolutions around the secondary body for the Titan DRO transfer, compared to $10.3$ revolutions for the Europa DRO transfer. 
This difference arises from the larger energy gap of $4.11\times10^{-3}$ NU between departure and arrival orbits in the Saturn-Titan system, compared to  $2.38\times10^{-3}$ NU for the Jupiter-Europa system.
The larger number of revolutions makes the Titan DRO more sensitive to small variations in the initial costates, leading to a highly nonconvex objective landscape with numerous local minima, small basins of attraction, and steep gradients. 
As shown in our previous study \cite{graebner2025aas}, directly targeting the Titan DRO transfer with baseline samples from the Europa DRO transfer makes it challenging to generate Pareto-optimal solutions. 

We therefore introduce a dimensionless homotopy parameter \(h\) as
\begin{equation}
h(\mu)=\frac{\mu-\mu_{\text{JE}}}{\mu_{\text{ST}}-\mu_{\text{JE}}},
\quad \mu_{\text{JE}}=2.525\times10^{-5},\;
\mu_{\text{ST}}=2.366\times10^{-4},
\end{equation}
so that \(h=0\) corresponds to the Jupiter–Europa system and \(h=1\) to the Saturn–Titan system.
Successively increasing $h$ in intermediate steps from the Europa DRO transfer to the Titan DRO transfer accelerates convergence of the Markov chain at each step to the intermediate target distributions. 
At each intermediate $h$-value, we create an artificial CR3BP system with natural units interpolated linearly between the boundary cases. 
Thruster parameters are kept constant in the natural units of each system, and 
corresponding closed departure and arrival DROs are determined at each step.   

\subsection{MALA Performance}
\begin{table}[t!]
  \centering
  \caption{Input parameters for the different MCMC algorithms. 
  Values in brackets indicate adapted parameters for the final iterations with smaller stepsizes.}
  \label{tab:mcmc_params_homotopy_adapted}
  \begin{tabular}{@{}lccc@{}}
    \toprule
    \textbf{Input Parameters} & \textbf{RWM} & \textbf{MALA} & \textbf{HMC} \\
    \midrule
    Objective scaling $\kappa_1$ 
      & $1.0$ 
      & $1.2$ 
      & $1.2$ \\

    Shooting time scaling $\kappa_2$ 
      & $1\times10^{-6}$ 
      & $1\times10^{-6}$ 
      & $1\times10^{-5}$ \\

    Proposal std.\ devs.\ $\boldsymbol{\sigma}_{\lambda}/\boldsymbol{\sigma}_{\text{init}}$ 
      & $0.05$ ($0.002$)
      & $0.02$ ($0.005$)
      & $0.006$ ($0.005$) \\

    Gradient timestep $\epsilon$ 
      & -
      & $2.5$ ($0.1$)
      & $0.5$ ($0.1$) \\

    Leapfrog steps $L$ 
      & -
      & -
      & $3$ ($1$) \\
      
    Scaling factor $\beta$ 
      & $10{,}000$ ($200{,}000$)
      & $10{,}000$ ($200{,}000$)
      & $10{,}000$ ($200{,}000$) \\

    Number of chains $N$ 
      & $1{,}920$ 
      & $1{,}920$ 
      & $1{,}920$ \\

    Number of iterations $M$ 
      & $3{,}000$ 
      & $350$ 
      & $140$ \\

    Burn-in iterations $M_0$ 
      & $2{,}470$ 
      & $270$ 
      & $105$ \\
    \bottomrule
  \end{tabular}
\end{table}
Starting with 1,920 samples from the baseline diffusion model, we run MALA for a total of $517$ hours across $96$ CPUs.
All algorithmic parameters are displayed in Table \ref{tab:mcmc_params_homotopy_adapted}.
The MALA parameters are mostly chosen based on those used for the random-walk Metropolis (RWM) algorithm in our previous work \cite{graebner2025aas}, shown in the adjacent column.
Different values of the objective function scaling parameters  $\kappa_1$ and $\kappa_2$ are selected empirically, as they target the Pareto front slightly better for this algorithm. 
Specifically, it is possible to increase the $\Delta v$-minimization component from $\kappa=1.0$ to $\kappa=1.2$ without a decrease in feasibility, as was previously observed for the RWM algorithm.
As before, the proposal covariance $\boldsymbol{\Sigma}_{\lambda}=\text{Diag}(\boldsymbol{\sigma}_{\lambda}^2)$ is chosen based on the empirical standard deviation of the initial samples $\boldsymbol{\sigma}_{\text{init}}$.
Here $\boldsymbol{\sigma}_{\lambda}$ is a vector consisting of the  standard deviation $\sigma_{\lambda,i}$ for each direction $i$ and $\boldsymbol{\sigma}_{\text{init}}=[0.0468,0.0010,0.0013,0.0353]$.
The standard deviation scaling factor $\boldsymbol{\sigma}_{\lambda}/\boldsymbol{\sigma}_{\text{init}}$ is chosen smaller (0.02 instead of 0.05) to avoid increasing the total step size when including the gradient drift term. 
\begin{figure}[b!]
\centering\includegraphics[width=\textwidth]{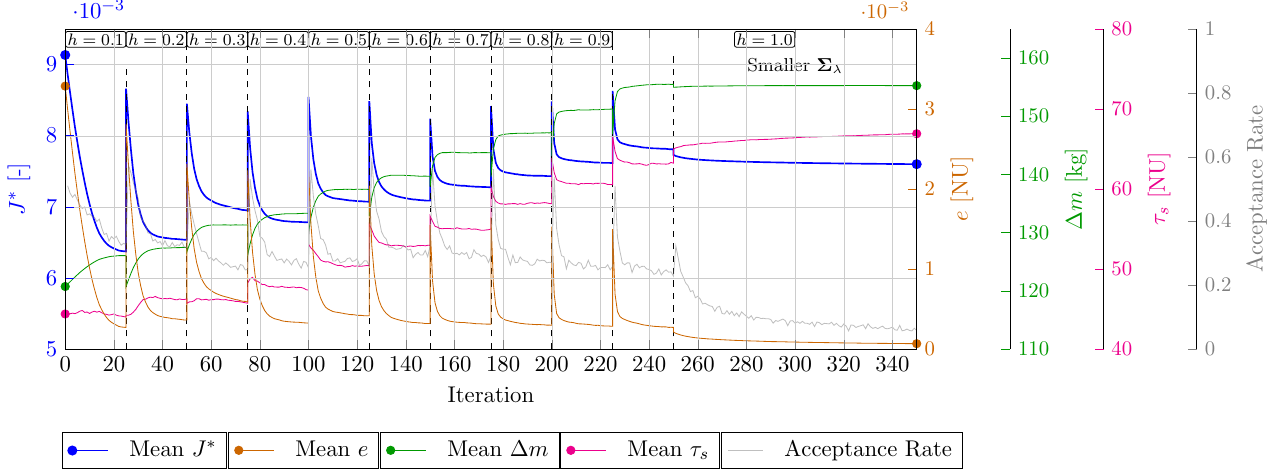}
	\caption{Mean values of the objective function and its three components during MALA.}
	\label{fig:MALA_iter_plot}
\end{figure}
The gradient timestep $\epsilon$ is chosen such that the gradient step size is, on average, approximately equal to the size of the random step. 
Empirically, this value achieves the best performance, with proposal quality degrading for larger values.

MALA proposes high-quality samples that rapidly decrease the mean objective value across all chains. 
This is visualized in Figure \ref{fig:MALA_iter_plot}, which shows the mean values of the objective and its three components across all chains over the iterations. 
Each homotopy step consists of $25$ iterations, followed by an additional $100$ iterations with smaller stepsize in the final stage, resulting in a total of $350$ iterations.
These final iterations further reduce the mean constraint violations by more effectively capturing local minima through a smaller proposal covariance and gradient step (values shown in brackets in Table \ref{tab:mcmc_params_homotopy_adapted}).
MALA proposes high quality samples, which quickly decrease the mean objective value across all chains. This is visualized in Figure \ref{fig:MALA_iter_plot}, which shows the mean values of the objective and its three components across all chains over the iterations. 
Each homotopy step consists of $25$ iterations, with an additional $100$ iterations with smaller stepsize in the final stage, leading to a total of $350$ iterations.
These final steps allow to further drive down the mean constraint violations by better capturing local minima through smaller proposal covariance and gradient steps (values shown in brackets in Table \ref{tab:mcmc_params_homotopy_adapted}).
When the algorithm switches to the next system in the homotopy scheme, the objective value initially increases and then quickly decreases. The mean fuel consumption also increases at the beginning of each homotopy stage, as a larger $h$ requires a higher $\Delta v$ to achieve feasibility. 
After this initial rise, the mean fuel consumption remains mostly constant.

\subsection{HMC Performance} 
We test the HMC algorithm on the same problem, starting with the same 1,920 samples from the baseline distribution. 
All algorithmic parameters are provided in the right column of Table \ref{tab:mcmc_params_homotopy_adapted}.
Because each proposal includes $L$ leapfrog integration steps, each iteration is computationally more expensive than in MALA.
Running the algorithm for a total of $140$ iterations takes $604$ hours, distributed across $96$ CPUs.
The number of leapfrog integration steps is kept relatively small at $L=3$ to avoid excessive computational cost per iteration. 
For this problem, increasing the total number of iterations provides a better return than allocating the same effort to further increasing $L$, since raising $L$ directly reduces how many iterations can be afforded for a fixed computational budget.
All other parameters are chosen to be similar to the MALA algorithm, with slight modifications chosen empirically based on improved performance.

\begin{figure}[t!]
\centering\includegraphics[width=\textwidth]{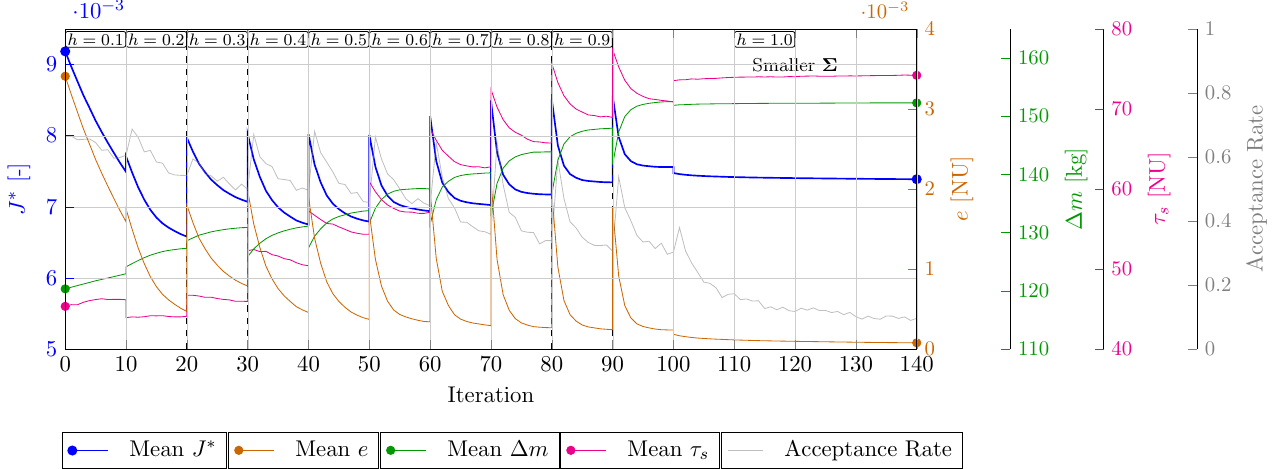}
	\caption{Mean values of the objective function and its three components during HMC.}
	\label{fig:HMC_iter_plot}
\end{figure}

The evolution of the objective function over the iterations is comparable to the MALA case, as shown in Figure \ref{fig:HMC_iter_plot}.
While only $10$ iterations per homotopy stage are not enough to reduce the mean objective to a comparable level achieved in the MALA case with $25$ iterations for $h=0.1$, the subsequent homotopy stages compensate for this, resulting in a final state with mean constraint violation below $1\times10^{-4}$.
Figure \ref{fig:HMC_iter_plot} also shows that the acceptance rate is consistently higher for HMC than for MALA (Figure \ref{fig:MALA_iter_plot}).
This is expected, as the leapfrog integration is designed to follow regions of high target density, thereby proposing higher-quality samples that are more likely to be accepted.
In the final stage an additional $40$ iterations with decreased stepsize are added (parameters in brackets in Table \ref{tab:mcmc_params_homotopy_adapted}).
Because the goal in this final stage is solely to better resolve local minima, rather than to discover new ones, we set $L=1$, effectively reducing the method to MALA.

\subsection{Comparison}
We compare the two presented gradient-based MCMC algorithms to the RWM algorithm employed in our previous work \cite{graebner2025aas}.
Multiple performance metrics, as well as computational cost, are considered in the comparison.
The results show that incorporating gradient information leads to overall improved performance, even when accounting for the additional computational overhead. 
\begin{table}[b!]
  \centering
  \captionof{table}{Comparison of performance and computational cost for the three MCMC algorithms. Mean and standard deviation (STD) are reported for both objectives; feasibility rate is based on $e<5\times10^{-5}$. All metrics are based on samples collected after the burn-in phase ($15,000$ to $25,000$ per method).}
  \label{tab:MCMC_results}
  \begin{tabular}{@{}lccc@{}}
    \toprule
    & \textbf{RWM} & \textbf{MALA} & \textbf{HMC} \\
    \midrule

    \multicolumn{4}{l}{\textbf{Performance}} \\
    \cmidrule(lr){1-4}
    $\Delta v$ (Mean $\pm$ STD) [m/s]  & $185.65\pm2.97$ & $184.95\pm4.15$ & $181.86\pm2.59$ \\
    $\tau_s$ (Mean $\pm$ STD) [days]   & $149.59\pm13.82$ & $158.56\pm24.28$ & $170.83\pm18.41$ \\
    Feasibility rate                   & 17.34\,\%        & 63.01\,\%        & 37.45\,\% \\[3pt]

    \multicolumn{4}{l}{\textbf{Computational Cost}} \\
    \cmidrule(lr){1-4}
    Runtime [CPU hours]                & $357$ & $517$ & $604$ \\
    Num.\ of function evaluations    & $5,760,000$ & $672,000$ & $652,800$ \\
    Num.\ of gradient evaluations      & $0$ & $672,000$ & $652,800$ \\
    \bottomrule
  \end{tabular}
\end{table}

\begin{figure}[t!]
  \centering
  \newlength{\triFigH}
  \setlength{\triFigH}{5.0cm} 

  \begin{minipage}[c]{0.37\textwidth}
    \centering
    \includegraphics[height=\triFigH,keepaspectratio]{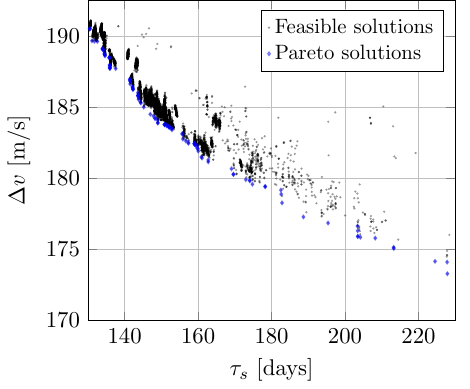}
  \end{minipage}\hfill
  \begin{minipage}[c]{0.31\textwidth}
    \centering
    \includegraphics[height=\triFigH,keepaspectratio]{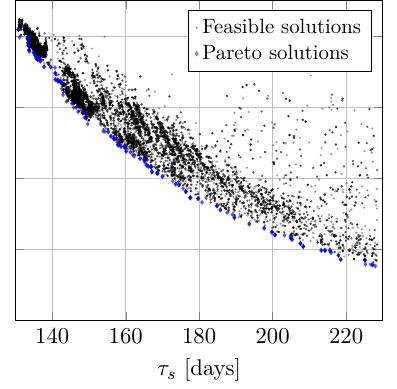}
  \end{minipage}\hfill
  \begin{minipage}[c]{0.31\textwidth}
    \centering
\includegraphics[height=\triFigH,keepaspectratio]{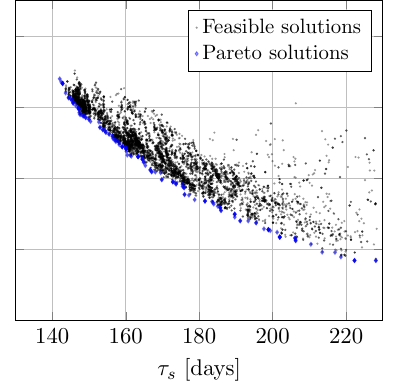}
  \end{minipage}

  \caption{Feasible samples (tolerance $e<5\times10^{-5}$ NU) from RWM (left), MALA (middle) and HMC(right), shown in the $\Delta v$-$\tau_s$ plane.}
  \label{fig:dv-TOF_saturn_comp}
\end{figure}

Feasibility is measured based on the percentage of samples with constraint violations $e<5\times10^{-5}$.
Table \ref{tab:MCMC_results} shows that MALA achieves by far the highest feasibility, almost quadrupling the RWM value from $17.34\,\%$ to $63.01\,\%$.
While HMC also increases the feasibility rate relative to RWM, it does not reach the MALA level, likely due to the smaller number of iterations.
Mean values of $\Delta v$ and shooting time $\tau_s$ for all feasible samples indicate that MALA achieves a lower $\Delta v$ than RWM, at the cost of a larger shooting time. 
HMC shows an even stronger trend towards lower $\Delta v$ and larger $\tau_s$.
This can also be seen in Figure \ref{fig:dv-TOF_saturn_comp}, which shows all feasible samples for both methods in the space of the two computing objectives. 
The figure demonstrates that HMC misses part of the Pareto front at smaller $\tau_s$.
Even though the larger shooting time scaling factor for HMC ($1\times10^{-5}$ instead of $1\times10^{-6}$) helps target the region of low $\tau_s$, the algorithm still excludes portions of that region.
Overall, the gradient-based methods achieve a denser Pareto front, especially for larger $\tau_s$.
The MALA samples also exhibit the greatest diversity in the objective space, observed both visually and quantitatively through the larger standard deviations reported in Table \ref{tab:MCMC_results}.

Computationally, the per-iteration cost is significantly higher for the gradient-based methods, with HMC being the most expensive. 
Computing the gradient involves numerically propagating the STM, which consists of quadratically more variables than the state. 
Due to the high sensitivity of the problem, this propagation must be highly accurate and requires precise interpolation of the switching times.
Therefore, even though MALA achieves the presented results with a significantly lower number of iterations and function evaluations, its runtime is still higher than that of RWM.
However, given the substantially better results, MALA remains the favorable approach.
A detailed study of how decreasing the integration and interpolation accuracy affects the gradient accuracy could improve the computational cost of gradient-based methods. 
Each HMC iteration is even more expensive, as it includes $L$ gradient evaluations per iteration, resulting in the longest overall runtime. 
For this problem, the additional cost is not justified in relative to MALA.
However, for more complex problems where the initial distribution differs substantially from the target distribution, this conclusion may change.

\subsection{DM Fine-tuning}
As the final step of our framework, the high-quality samples generated by MALA, together with their reward values, are used to fine-tune the baseline diffusion model according to Eq. \eqref{eq:updated loss}.
Of the total $21,182$ samples generated by the algorithm, the worst $10\,\%$ in terms of objective value $J^*$ are discarded, and the remaining samples are normalized and used for training. 
The diffusion model architecture and parameters are identical to those in our previous work, which provides a more detailed description of this procedure \cite{Graebner.1132025}.
Using a GPU, model training takes approximately $15$ minutes, and the generation of $50{,}000$ samples takes about $13$ minutes.

\begin{figure}[t!]
  \centering
  \setlength{\triFigH}{5.0cm} 

  \begin{minipage}[c]{0.31\textwidth}
    \centering
    \includegraphics[height=\triFigH,keepaspectratio]{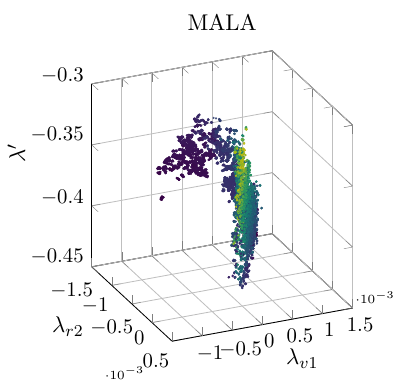}
  \end{minipage}\hfill
  \begin{minipage}[c]{0.31\textwidth}
    \centering
    \includegraphics[height=\triFigH,keepaspectratio]{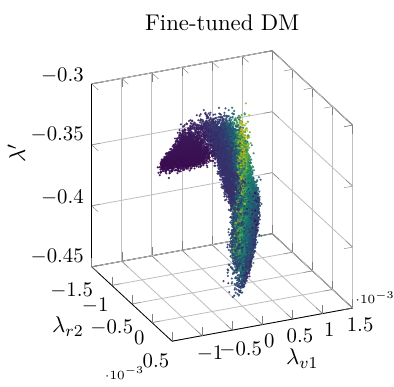}
  \end{minipage}\hfill
  \begin{minipage}[c]{0.37\textwidth}
    \centering
\includegraphics[height=\triFigH,keepaspectratio]{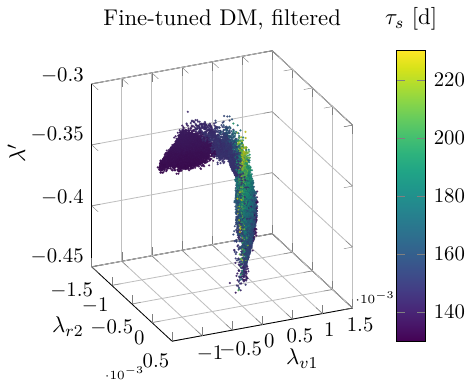}
  \end{minipage}
  \caption{The $19,000$ final MALA samples (left), $50,000$ samples from the fine-tuned DM (middle) and a subset of $21,000$ feasible DM samples (right) with $e<5\times10^{-5}$ displayed in the costate space.}
  \label{fig:costate_saturn_comp}
\end{figure}
Figure \ref{fig:costate_saturn_comp} presents a three-dimensional projection of the four-dimensional costate space by combining two elements of the costate vector:
\begin{equation}
    \lambda'=\cos{(0.7766)}\lambda_{r1}+\sin{(0.7766)}\lambda_{v2}.
\end{equation}
This transformed costate variable is introduced due to a strong linear relationship between $\lambda_{r1}$ and $\lambda_{v2}$ with slope $\tan (0.7766)$.
A similar relationship with a different slope for the related Europa DRO transfer is shown and explained in our previous work \cite{Graebner.1132025}.

Based on the limited number of samples, the diffusion model is able to learn the underlying data distribution, which enables it to generate new solutions.
The final MALA samples, shown on the left of Figure \ref{fig:costate_saturn_comp}, are approximately distributed according to the target distribution but exhibit some gaps.
The middle and right plots of Figure \ref{fig:costate_saturn_comp} show samples generated by the fine-tuned diffusion model and the feasible subset of those samples in the costate space.
This highlights the two major motivations for including the final diffusion model fine-tuning step in the framework. 
First, by learning an approximation $p_{\alpha}$ of the target distribution $\pi_{\alpha}$, we can generate an unlimited number of high-quality samples. 
Second, this process reveals new solutions, effectively filling gaps in the solutions space that were not reached by MALA within the given number of iterations.

\section{Conclusion}
This work investigates gradient-based MCMC algorithms for generating low-thrust spacecraft trajectories in the CR3BP as part of a self-supervised diffusion model fine-tuning framework.
Objective values are computed efficiently through a preliminary screening algorithm, and gradients are approximated through a fixed-time objective, enabling analytic evaluation via STMs.
Compared to RWM, both MALA and HMC significantly increase the feasibility rate and produce denser coverage of the Pareto front in the competing objectives $\Delta v$ and shooting time. 
Among the tested methods, MALA delivers the strongest overall performance, nearly quadrupling the number of feasible solutions relative to RWM while incurring only modest additional runtime. 
HMC exhibits similar advantages but falls short of MALA, primarily due to its higher per-iteration cost and the resulting limitation on the number of realizable iterations under a fixed computational budget. 
Although gradient evaluations introduce additional computational overhead, they consistently reduce the total number of required iterations by improving the proposal quality.
As a final step in the workflow, fine-tuning a diffusion model on high-quality MALA samples provides a scalable means of approximating the target distribution. 
This enables the generation of large numbers of new solutions and effectively fills gaps in the solution space that were not reached by MCMC alone.

Overall, the results demonstrate that combining gradient-based sampling with diffusion-model fine-tuning constitutes an effective and extensible framework for generating Pareto-optimal trajectories in the CR3BP. 
Future work will explore extending these methods to more complex mission designs, conditioning on additional problem parameters, and further reducing computational costs.

\section{Acknowledgement}
Simulations were performed on computational resources managed and supported by Princeton Research Computing, a consortium of groups including the Princeton Institute for Computational Science and Engineering (PICSciE) and the Office of Information Technology’s High-Performance Computing Center and Visualization Laboratory at Princeton University.
We also acknowledge partial travel support from the Princeton School of Engineering and Applied Science (SEAS) through a SEAS Travel Award.
\bibliography{references}

\end{document}